\documentclass[11pt]{article}
\usepackage{amssymb,amsmath,amsthm}
\usepackage{authblk}
\usepackage{multirow}
\usepackage{blindtext}
\usepackage{amsfonts}
\usepackage{slashed}
\usepackage{graphicx}
\usepackage{natbib}
\usepackage[libertine]{newtxmath}

\numberwithin{equation}{section}

\title{Nonparametric Pricing and Hedging of Volatility Swaps in Stochastic Volatility Models}
\author{Frido Rolloos\thanks{frolloos@yahoo.com}}

\date{March 28, 2020}

\begin{document}
\maketitle

\abstract{In this paper the zero vanna implied volatility approximation for the price of freshly minted volatility swaps is generalised to seasoned volatility swaps. We also derive how volatility swaps can be hedged using a strip of vanilla options with weights that are directly related to trading intuition. Additionally, we derive first and second order hedges for volatility swaps using only variance swaps. As dynamically trading variance swaps is in general cheaper and operationally less cumbersome compared to dynamically rebalancing a continuous strip of options, our result makes the hedging of volatility swaps both practically feasible and robust. Within the class of stochastic volatility models our pricing and hedging results are model-independent and can be implemented at almost no computational cost.}

\newpage

\section{Assumptions and notations}

We will work under the premise that the market implied volatility surface is generated by the following general stochastic volatility (SV) model
\begin{align}
\label{dS}
dS &=  \sigma S \left[ \rho \, dW + \bar{\rho} \, dZ  \right]  \\
\label{dsigma}
d\sigma &= a(\sigma,t) \, dt + b(\sigma, t) \, dW
\end{align}
where $\bar{\rho} = \sqrt{1-\rho^2}$, $dW$ and $dZ$ are independent standard Brownian motions, and the functions $a$ and $b$ are deterministic functions of time and volatility. The results derived in this paper are valid for any SV model satisfying \eqref{dS} and \eqref{dsigma}, which includes among others the Heston model, the lognormal SABR model, and the $3/2$ model. 

\smallskip
The SV process is assumed to be well-behaved in the sense that vanilla options prices are risk-neutral expectations of the payoff function:
\begin{equation}
C (S,K )= E_t \left[ \left( S(T) - K \right)_+ \right]
\end{equation}
The option price $C$ can always be expressed in terms of the Black-Merton-Scholes (BS) price $C^{BS}$ with an implied volatility parameter $I$:
\begin{equation}
C(S, K) = C^{BS} (S,K,I) 
\end{equation}
It is assumed that the implied volatility parameter $I = I(S,t,K,T,\sigma,\rho)$. The BS price is given by the following well-known formula,
\begin{equation}
C^{BS} (S,K,I) = S N(d_+) - K N(d_-)
\end{equation}
where $ N \left(d_{\pm} \right)$ are normal distribution functions, and
\begin{equation}
d_- = \frac{\log (S/K)}{ I \sqrt{\tau}} - \frac{I \sqrt{\tau}}{2}, \quad
d_+ = d_- + I \sqrt{\tau} 
\end{equation}
and $\tau = T-t$.

\smallskip
The generalised Hull-White formula also gives a relationship between options prices and BS prices
\begin{equation}
C(S,K) = E \left [ C^{BS} \left( S M_{t,T}(\rho),K, \sigma_{t,T} (\rho) \right) \right]
\end{equation}
where
\begin{equation}
M_{t,T}(\rho) = \exp \left\{ -\frac{\rho^2}{2}\int_t^T \sigma^2 \, du + \rho \int_t^T \sigma \, dW \right\}
\end{equation}
and $\sigma_{t,T} (\rho) $ is 
\begin{equation}\label{rvol}
\sigma_{t,T} (\rho) = \sqrt { \frac{1}{\tau}\int_t^T \bar{\rho}^2 \sigma^2 \, du}
\end{equation}
This can be written as
\begin{equation}\label{mf2}
C^{BS}(S,K,I) = E_t \left[ C^{BS} \left( SM_{t,T}(\rho),K, \sigma_{t,T}  (\rho) \right) \right]
\end{equation}

\smallskip
We will also need BS greeks for vanilla options, in particular the delta $(\Delta^{BS})$, vega $(\nu^{BS})$, vanna $(va^{BS})$, and volga $(vo^{BS})$:
\begin{align}
\Delta^{BS} (S, K,I)&= N(d_+) \\
\nu^{BS} (S,K,I) &= S \sqrt{\tau} \, N'(d_+)  \\
va^{BS} (S,K,I) &= - \frac{N'(d_+)}{I} \, d_- \\
vo^{BS} (S, K,I) &= \frac{S \sqrt{\tau} \, N'(d_+)}{I} \, d_+ d_-
\end{align}
Special notation will be given to the strike and corresponding implied volatility where $d_- = 0$:
\begin{equation}
\log S / K_- = \frac{1}{2} I^2_- \sqrt{\tau}
\end{equation}
$K_-$ is called the zero vanna strike, and $I_-$ the zero vanna implied volatility.

\smallskip
In \cite{RA} it is proved that for general SV models the price of a volatility swap at trade inception can be read directly from the market smile of European vanilla options, and is approximately equal to the zero vanna implied volatility. The approximation is accurate and valid for a large set of SV models, independent of the specific model parameters. Moreover, it has been demonstrated in \cite{ARS} that for SV models driven by fractional noise, the zero vanna implied volatility approximation remains robust. 

\smallskip
Except for at trade inception, however, a volatility swap is `seasoned', meaning that it will have a realised volatility component. To be able to price a volatility swap throughout its life we must therefore be able to calculate
\begin{equation}
\label{svs}
\mathcal{V}_{0,T}(t) = E_t \left[ \sqrt { \frac{1}{T}\int_0^T \sigma^2 \, du} \, \right]
\end{equation}
for all $t \in [0,T]$. Furthermore, if we are not able to price seasoned volatility swaps we will not know how to hedge volatility swaps as the change in the value of a volatility swap is the change in the seasoned volatility swap price.

\smallskip
To introduce notation which will later be used in our discussion on hedging of volatility swaps, the price of a variance swap is given the notation
\begin{equation}\label{vas}
V^2_{0,T}(t) = E_t \left[  \frac{1}{T} \int_0^t \sigma^2 \, du + \frac{1}{T} \int_t^T \sigma^2 \, du   \right]
\end{equation}
Variance swaps are clearly far easier to price and hedge compared to volatility swaps. In some markets, trading variance swaps is in fact more liquid than trading the theoretical replicating portfolio for variance swaps, which is a continuous strip of options. From that perspective it would be attractive to be able to hedge volatility swaps directly with variance swaps, and if possible in an as model-independent manner as possible.

\section{The historical adjusted spot process} 

\smallskip
Let us introduce an auxiliary geometric Brownian motion $H$ with constant volatility $c$,
\begin{equation}
\label{dH}
dH = c H \, dB
\end{equation}
In addition, we require that the process $H$ is independent of $S$ and $\sigma$, which means that $dB dW = dB dZ = 0$. We shall later see that $c$ is related to the historical realised volatility of $S$. 

\smallskip
Define the `historical adjusted' spot process as
\begin{equation}
\label{SH0}
S^H = S H
\end{equation}
The SDE for $S^H$ reads
\begin{equation}
\label{dSH0}
dS^H = S^H \left[ c \, dB + \rho \sigma \, dW + \bar{\rho} \sigma \, dZ \right]
\end{equation}
Equation \eqref{dSH0} can be rewritten in the following two equivalent ways:
\begin{align}
\label{dSH1}
dS^H &= S^H \left[ \sqrt{ c^2 + \bar{\rho}^2 \sigma^2} \, dW^\bot + \rho \sigma \, dW \right] \\
\label{dSH2}
dS^H &= S^H \left[ c \, dB + \sigma \, dB^\bot \right]
\end{align}
where
\begin{equation}
dW^\bot  = \frac{c \, dB + \bar{\rho} \sigma \, dZ}{\sqrt{ c^2 + \bar{\rho}^2 \sigma^2}}, \quad
dB^\bot  = \frac{\rho \sigma \, dW + \bar{\rho} \sigma \, dZ}{\sigma}
\end{equation}

\smallskip
Integrating the SDE \eqref{dSH1} gives
\begin{equation}
\label{SH1}
S^H_T = \widetilde{S}^H_T M_{t,T}(\rho)
\end{equation}
with
\begin{equation}
\widetilde{S}^H_T = S^H \exp \left \{  -\frac{1}{2} \int_t^T \left[ c^2 + \bar{\rho}^2 \sigma^2 \right] du + \int_t^T \sqrt{ c^2 + \bar{\rho}^2 \sigma^2} \, dW^\bot  \right \}
\end{equation}
\and
\begin{equation}
M_{t,T}(\rho) = \exp \left \{  -\frac{1}{2} \int_t^T \rho^2 \sigma^2 \, du + \int_t^T \rho \sigma \, dW    \right \}
\end{equation}

\smallskip
Similarly we can integrate \eqref{dSH2} to obtain
\begin{equation}
\label{SH2}
S^H_T = S_T H_T
\end{equation}
with
\begin{equation}
S_T = S \exp \left \{  -\frac{1}{2} \int_t^T  \sigma^2  \, du + \int_t^T \sigma \, dB^\bot  \right \}
\end{equation}
\and
\begin{equation}
H_T = H \exp \left \{  -\frac{1}{2} \int_t^T c^2 \, du+ \int_t^T c \, dB    \right \}
\end{equation}
In the remainder of the paper we will set the initial value of $H$ to one, i.e. $H=1$ and hence $S^H = S$.

\section{The historical adjusted volatility smile}

What we mean with the historical adjusted smile is the volatility smile $I^H$ of vanilla options on the process $S^H$. Although this process is not directly traded, we can express options on $S^H(T)$ in terms of options on $S$ by making use of \eqref{SH2} and the independence of $dB$ and $dB^\bot$. To construct $I^H$ note that by conditioning we can write
\begin{align}
E_t \left[ (S^H_T - K )_+ \right] &= E_t \left[ ( S_T H_T - K )_+ \right] \nonumber \\
&= E_t \left[ H_T C(S,K/H_T) \right] \nonumber \\
&= \int_0^\infty h C(S,K/h) q(h) dh
\end{align}
where $q(h)$ is the lognormal distribution because we have taken $c$ as constant:
\begin{equation}
q(h) = \frac{1}{h\nu\sqrt{2\pi}} \exp \left\{  - \frac{1}{2} \left(\frac{ \ln h- \mu}{\nu} \right)^2 \right\}, \, \nu = c\sqrt{\tau}, \, \mu =  - \frac{1}{2} c^2 \tau
\end{equation}
The options prices $C(S,K/h)$ under the integral are market prices of vanilla options on $S$ and are available for all strikes. Thus
\begin{equation}
C^H (S,K) = \int_0^\infty h C(S,K/h) q(h) \, dh
\end{equation}
where the $C^H$ left of the equality sign denotes the price of options on $S^H$ (recall we set $H=1$). In terms of BS prices we have the following equivalent expression:
\begin{equation}
\label{CBSH}
C^{BS} \left(S,K,I^H\right) = \int_0^\infty h C^{BS}(S,K/h,I) q(h) \, dh
\end{equation}

\smallskip
As all quantities on the right hand side of \eqref{CBSH} are known the numerical integration can be carried out to back out $I^H$. It is important to remember that for each $h$ the implied volatility $I$ on the right hand side of \eqref{CBSH} is the implied volatility corresponding to the strike $K/h$ and not the implied volatility corresponding to $K$.

\smallskip
We have priced options on $S^H_T$ making use of \eqref{SH2}. However, options on $S^H_T$ can also be priced using \eqref{SH1} and conditioning on $M_{t,T}(\rho)$. We shall see that equating the two naturally leads us to the fair strike of seasoned volatility swaps. Indeed, 
\begin{align}
E_t \left[ ( S^H_T - K )_+ \right] &= E_t \left[ ( \widetilde{S}^H_T M_{t,T}(\rho) - K )_+ \right] \nonumber \\
&= E_t \left[ C^{BS} \left( S M_{t,T}(\rho), K, \sigma^H_{t,T} (\rho) \right) \right]
\end{align}
with
\begin{equation}
\sigma^H_{t,T} (\rho) = \sqrt{ \frac{1}{\tau} \int_t^T \left[ c^2 + \bar{\rho}^2 \sigma^2 \right] du }
\end{equation}
Hence,
\begin{equation}
\label{mixing}
C^{BS} \left(S,K, I^H\right)  = E_t \left[ C^{BS} \left( S M_{t,T}(\rho), K, \sigma^H_{t,T} (\rho) \right) \right]
\end{equation}

\section{Volatility swaps pricing}

\smallskip
Following the method introduced in \cite{RA} the right hand side of \eqref{mixing} can be Taylor expanded around $\rho=0$ and around the historical adjusted implied volatility $I^H_-$ where the BS vanna of a vanilla option on $S^H_T$ is zero. The Taylor expansion of \eqref{mixing} around $\rho = 0$ gives
\begin{align}
C^{BS} (S,K, I^H) &\approx E_t \left[ C^{BS} ( S, K, \sigma^H_{t,T} (0) ) \right] \nonumber \\
&\quad + \rho S E_t \left[ \Delta^{BS} ( S, K, \sigma^H_{t,T} (0) ) \int_t^T \sigma \, dW  \right]
\end{align}
with
\begin{equation}
\sigma^H_{t,T} (0) = \sqrt{ \frac{1}{\tau} \int_t^T \left[ c^2 + \sigma^2 \right] du }
\end{equation}

\smallskip
Now, restricting to an option with zero vanna strike $K_-$ and zero vanna historical adjusted implied volatility $I^H_-(t)$, and expanding around $I^H_-(t)$, we obtain
\begin{align}
C^{BS} (S,K_-,I^H_- ) &\approx C^{BS} ( S, K_-, I^H_- ) \nonumber \\
&\quad + \nu^{BS} (S, K_-, I^H_-) \, E_t \left[ ( \sigma^H_{t,T} (0) - I^H_- ) \right] \nonumber \\
&\quad + \frac{1}{2} vo^{BS} ( S, K_-, I^H_- ) \, E_t \left[ ( \sigma^H_{t,T} (0) - I^H_- )^2 \right] \nonumber \\
&\quad + \rho S va^{BS} ( S, K_-, I^H_- ) \, E_t \left[ ( \sigma^H_{t,T} (0) - I^H_- ) \int_t^T \sigma \, dW  \right]
\end{align}
Since
\begin{equation}
 \nu^{BS} ( S, K_-, I^H_- ) = vo^{BS} ( S, K_-, I^H_- ) = 0
 \end{equation}
 this reduces to
\begin{equation}
C^{BS} (S,K_-, I^H_- ) \approx C^{BS} ( S, K_-,  I^H_- ) + \nu^{BS} ( S, K_-, I^H_-) \, E_t \left[ ( \sigma^H_{t,T} (0) - I^H_- (t) ) \right] 
\end{equation}
This can only be the case if
\begin{equation}
E_t \left[ \sigma^H_{t,T} (0) \right] = E_t \left[ \sqrt{ \frac{1}{\tau} \int_t^T \left[ c^2 + \sigma^2 \right] du } \, \right] \approx I^H_-
\end{equation}

\smallskip
Define the constant $c$ as
\begin{equation}
\label{defc}
c^2 = \frac{1}{\tau} \int_0^t \sigma^2 \, du
\end{equation}
and we obtain our desired result:
\begin{equation}\label{priceresult}
\mathcal{V}_{0,T}(t) = E_t \left[ \sqrt{ \frac{1}{T} \int_0^t \sigma^2 \, du + \frac{1}{T} \int_t^T \sigma^2 \, du } \, \right] \approx I^H_- \sqrt{ \frac{\tau}{T} }
\end{equation}
The above equation contains as a special case the formula for freshly minted volatility swaps derived in \cite{RA}.

\smallskip
We summarize the steps required to calculate the price of a seasoned volatility swap at time $t$ with a realized volatility given by \eqref{defc}: Given a market smile which is assumed to be generated by a process of the type \eqref{dS} - \eqref{dsigma}, introduce an adjusted spot process defined by \eqref{dH} - \eqref{SH0}. Options on the adjusted spot process of maturity corresponding to the maturity of the volatility swap can be priced with \eqref{CBSH}. Once options on the adjusted process have been priced the adjusted implied volatility can be backed out. The final step is to find the adjusted implied volatility where an option on the adjusted process has zero BS vanna. Equation \eqref{priceresult} then gives the approximate value of the seasoned volatility swap.

\smallskip
Note that at time $t' = t + dt$ we will update the value of the `constant' $c^2$ to $c'^2$ to take into account of the new historical realised volatility, and set $H(t') = 1$ again in order to calculate the updated seasoned volatility swap price. The aforementioned quantities should therefore be treated only as a tool to calculate and not as actual physical processes. What we mean with $c$ is constant in this context is that at time $t$ we treat it as if it will be constant until maturity date in order to be able get the historical volatility `under the square root sign'. In Appendix A we show that the pricing as described above gives the exact seasoned volatility swap price in a BS setting.

\section{Volatility swaps hedging}

\subsection{Hedging with options}

In order to find the hedge for a (seasoned) volatility swap suppose that at time $t$ we have found the historical adjusted zero vanna strike and implied volatility for a vanilla option on the historical adjusted spot price. If we buy this `meta option', which according to equation \eqref{CBSH} is a strip of market traded vanilla options, the change in value of the option over $dt$ is
\begin{equation}
\label{dCBSH}
 d C^{BS} (S,K_-,I^H_-) = \int_0^\infty h [dC^{BS} (S,K_-/h,I ) ] q(h) \, dh
 \end{equation}
where in the integrand $I = I(S, K_-/h) $. Hence,
\begin{equation}
E_t \left[  d C^{BS} (S,K,I^H) \right ] = 0
\end{equation}

\smallskip
Next, note that as  $C^{BS} (S,K_-,I^H_- )$ satisfies the BS partial differential equation, and taking into account that the vanna and volga contributions are zero,
\begin{equation}\label{BSpde}
\frac{ \sqrt{\tau} }{ \nu^{BS} (S, K_-,I^H_-) } \left[ d C^{BS} (S,K_-,I^H_-) - \Delta^{BS} (S, K_-,I^H_-) dS \right] = d ( I^H_- \sqrt{\tau} \, ) - \frac{1}{2} \frac{ \sigma^2 }{ I^H_- \sqrt{\tau} } \, d\tau  
\end{equation}
The risk-neutral drift of the zero vanna implied volatility at time $t$ is thus
\begin{equation}\label{driftzv}
E_t \left[ d ( I^H_- \sqrt{\tau} \, ) \right] = \frac{1}{2} \frac{ \sigma^2 }{ I^H_- \sqrt{\tau} } \, d \tau
\end{equation}

\smallskip
As the volatility swap price is a martingale, and from equation \eqref{priceresult} we see that at each instant it is (approximately) equal to the zero vanna implied volatility, the change in a fixed strike implied volatility which is initially zero vanna cannot be equal to the change in the volatility swap price:
\begin{equation}
d (\mathcal{V}_{0,T}(t) T ) \neq d ( I^H_- \sqrt{\tau} \, )
\end{equation}
This is because a fixed strike implied volatility which is zero vanna at time $t$ will not be zero vanna at $t+dt$. Indeed, from equation \eqref{driftzv} it is clear that an initially zero vanna implied volatility is not a martingale. The new zero vanna implied volatility at $t+dt$ will correspond to a different strike.

\smallskip
Let us introduce the \emph{total} change in zero vanna implied volatility $D ( I^H_- \sqrt{\tau} \, )$. This total change must satisfy
\begin{equation}\label{dvs=DI}
d (\mathcal{V}_{0,T}(t) T )  \approx D ( I^H_- \sqrt{\tau} \, )
\end{equation}
The total change, which is the change from the zero vanna implied volatility at $t$ to the zero vanna implied volatility at $t+dt$, can be written as:
\begin{equation}
\label{DI}
D ( I^H_- \sqrt{\tau} \, ) = d ( I^H_- \sqrt{\tau} \, ) + \delta ( I^H_- \sqrt{\tau} \, ) + \frac{ \partial ( I^H_- \sqrt{\tau} \, ) }{ \partial \log K} d\log K_- + \mathcal{H.O.}
\end{equation}
The term $d ( I^H_- \sqrt{\tau} \, )$ is the usual change in fixed strike implied volatility. However, we know that to recalculate the volatility swap price the new realised volatility needs to be taken into account. Since the infinitesimal change in realised volatility is $\sigma^2 \, dt$, the change in fixed strike implied volatility purely due to the realised volatility update will be of order $dt$: $\delta ( I^H_- \sqrt{\tau} \, ) = O(dt)$. The term involving the slope of the adjusted implied volatility skew at the initial zero vanna strike is the correction needed to find the new zero vanna implied volatility strike after the usual change in implied volatility and change due to the realised volatility update.

\smallskip
Recall that the zero vanna strike and implied volatility at $t$ satisfies
\begin{equation}
\log(S/K_-) = \frac{1}{2} ( I^H_- \sqrt{\tau} \, )^2
\end{equation}
We can therefore relate the total change in zero vanna adjused implied volatility to the required change in strike to maintain zero vanna:
\begin{equation}
d \log K_- = d \log S - I^H_- \sqrt{\tau} \, D ( I^H_- \sqrt{\tau} \, ) + \frac{1}{2} ( D ( I^H_- \sqrt{\tau} \,) )^2
\end{equation}
Substituting this into equation \eqref{DI} and after some rearranging, gives
\begin{align}\label{DI2}
\left( 1 + I^H_- \sqrt{\tau} \, \frac{ \partial (I^H_- \sqrt{\tau} \, ) }{ \partial \log K} \right) D ( I^H_- \sqrt{\tau} \, ) &= d ( I^H_-\sqrt{\tau} \, )  + \delta ( I^H_- \sqrt{\tau} \, )  + \frac{ \partial ( I^H_- \sqrt{\tau} \, ) }{ \partial \log K} d\log S \nonumber \\
&\quad + \mathcal{H.O.}
\end{align}

\smallskip
Using equation \eqref{dvs=DI}, it follows that 
\begin{equation}
 E_t \left[ D ( I^H_- \sqrt{\tau} \, ) \right]  \approx E_t \left[ d (\mathcal{V}_{0,T}(t) T ) \right] = 0
 \end{equation}
 and hence, because $\delta ( I^H_- \sqrt{\tau} \, ) = O(dt)$, it follows from \eqref{DI2} that
 \begin{align}
 \delta ( I^H_-(t) \sqrt{\tau} \, ) &= E_t \left[ \delta ( I^H_- \sqrt{\tau} \, ) \right ] \nonumber \\
 &\approx - E_t \left[ d ( I^H_- \sqrt{\tau} \, ) \right] - \frac{ \partial ( I^H_- \sqrt{\tau} \, ) }{ \partial \log K} E_t \left[ d\log S \right] - \mathcal{H.O.} \nonumber \\
 &\approx -\frac{1}{2} \frac{ \sigma^2 }{ I^H_- \sqrt{\tau} } \, d \tau - \frac{ \partial ( I^H_- \sqrt{\tau} \, ) }{ \partial \log K} E_t \left[ d\log S \right] - \mathcal{H.O.}
\end{align}
Substituting this back into \eqref{DI2},
\begin{equation}\label{DI3}
\left( 1 + I^H_- \sqrt{\tau} \, \frac{ \partial ( I^H_- \sqrt{\tau} \, ) }{ \partial \log K} \right) D ( I^H_- \sqrt{\tau} \, ) \approx d ( I^H_- \sqrt{\tau} \, )  - \frac{1}{2} \frac{ \sigma^2 }{ I^H_- \sqrt{\tau} } \, d \tau + \frac{ \partial ( I^H_- \sqrt{\tau} \, ) }{ \partial \log K} \, \frac{dS}{S} 
\end{equation} 
because $d \log S = E_t [d \log S ] + dS / S$. The above can be simplified further by noting that for the class of SV models considered,
\begin{equation}
\frac{ \partial ( I^H_- \sqrt{\tau} \, ) }{ \partial \log K} = - \frac{ \partial ( I^H_- \sqrt{\tau} \, ) }{ \partial \log S}
\end{equation}
and so,
\begin{equation}\label{DI3}
\left( 1 + I^H_- \sqrt{\tau} \, \frac{ \partial ( I^H_- \sqrt{\tau} \, ) }{ \partial \log K} \right) D ( I^H_- \sqrt{\tau} \, ) \approx d ( I^H_- \sqrt{\tau} \, )  - \frac{1}{2} \frac{ \sigma^2 }{ I^H_- \sqrt{\tau} } \, d \tau - \frac{ \partial ( I^H_- \sqrt{\tau} \, ) }{ \partial S} \, dS 
\end{equation}

\smallskip
Comparing equation \eqref{DI3} with the BS PDE \eqref{BSpde} we see that
\begin{equation}\label{hedgeresult}
d (\mathcal{V}_{0,T}(t) T ) \approx D ( I^H_- \sqrt{\tau} \, ) \approx N \left[ d C^H ( S,K_- ) -  \Delta^H (S, K_-) \, dS \right]
\end{equation}
where $\Delta^H$ is the SV skew-adjusted delta
\begin{equation}\label{hedgedelta}
\Delta^H (S, K_-) = \Delta^{BS} (S, K_- , I^H_-) + \nu^{BS} (S, K_-,I^H_-) \, \frac{ \partial  I^H_-  }{ \partial S}
\end{equation}
and the notional $N$ is
\begin{equation}\label{hedgenotional}
N = \left( \frac{  \nu^{BS} (S, K_-,I^H_-) }{ \sqrt{\tau} } \right)^{-1} \left( 1 + I^H_- \sqrt{\tau} \, \frac{ \partial ( I^H_- \sqrt{\tau} \, ) }{ \partial \log K} \right)^{-1}
\end{equation}

\smallskip
What the hedging formula \eqref{hedgeresult} says is that the change in volatility swap price can be approximately hedged in a self-financing manner by trading a delta-hedged zero vanna option on the historical adjusted spot process, where the delta is the skew adjusted delta, and the notional is a skew adjusted and vega-weighted notional. Note that this formula is identical in form to the formula in [] for hedging forward starting volatility swaps. The reason that there is no delta hedge in the formula for forward starting volatility swaps is because the forward starting options used in the paper are insensitive to the spot price movements.

\smallskip
As in \cite{CL}, our hedge for the volatility swap involves continuous rebalancing of vanilla options of all strikes with appropriate weights (recall that a vanilla option on the adjusted spot process is a strip of market traded vanilla options). The difference between our approach and \cite{CL} is that the weights are directly related to intuitive trading concepts such as BS greeks and slope of the skew. Additionally, the weights are smooth functions of strike and not highly oscillatory.

\smallskip
Even though hedging with theoretical strip of options is, within our approach and framework, the most accurate hedge for the volatility swap, it is not the cheapest way to hedge. For this reason, the next section will discuss an approach to hedge volatility swaps using variance swaps only, which for some indices is cheaper to trade than a portfolio consisting of options of all possible strikes. The price to pay for hedging with relatively cheaper instruments is potential loss of accuracy (increased hedging error). 

\subsection{Hedging with variance swaps}

\subsubsection{Gatheral's formula}

Two approximations for hedging volatility swaps with variance swaps will be given: a first order and second order approximation. Both will be based on Gatheral's formula for variance swaps. We recall that for a freshly minted variance swap at time $t$ it reads
\begin{equation}
V^2_{t,T} (t) \tau = \int_{-\infty}^\infty  N'(d_- ) \frac{ \partial d_- }{\partial \log K} \, I^2 \tau \, d \log K
\end{equation}
This formula can be generalised to seasoned volatility swaps. Note that the price of a seasoned volatility swap is given by
\begin{equation}
V^2_{0,T} (t) T = -2 E_t \left[ \log S^H_T/ S^H \right]
\end{equation}
with $c$ given by \eqref{defc}. A log-contract on the adjusted spot process $S^H$ can be synthesised using options on the adjusted spot process $C^{BS}(S,K,I^H)$. Hence, Gatheral's formula for seasoned volatility swaps reads
\begin{equation}
V^2_{0,T} (t) T = \int_{-\infty}^\infty  N'(d_- ) \frac{ \partial d_- }{\partial \log K} \, (I^H \sqrt\tau)^2 \, d \log K
\end{equation}
which can be rewritten in the slightly more convenient form
\begin{equation}\label{Gath}
V^2_{0,T} (t) T = \int_{-\infty}^\infty  N'(d_- ) (I^H \sqrt\tau)^2 \, d d_-
\end{equation}

\subsubsection{First order approximation}

The first order approximation is found by expanding $I^H \sqrt\tau$ around $d_- =0$. This leads to
\begin{align}
\int_{-\infty}^\infty  N'(d_- ) (I^H \sqrt\tau)^2 \, d d_- &\approx (I^H_- \sqrt\tau)^2 \int_{-\infty}^\infty  N'(d_- ) \, d d_- +  \frac{ \partial (I^H_- \sqrt\tau)^2 }{\partial d_-} \int_{-\infty}^\infty  d_- N'(d_- ) \, d d_- + \nonumber \\
&=  (I^H_- \sqrt\tau)^2
\end{align}
Hence, the lowest order approximation is
\begin{equation}
V^2_{0,T} (t) T \approx (I^H_- \sqrt\tau)^2 \approx \mathcal{V}^2_{0,T} (t) T
\end{equation}
From which it follows that
\begin{equation}\label{firstorder}
d \mathcal{V} _{0,T} (t) \approx \frac{1} {2 I^H_- \sqrt\tau} d V^2_{0,T} (t)
\end{equation}
By dynamically trading variance swaps with notional $1 / (2 I^H_- \sqrt\tau)$ we can hedge a volatility swap. Note however that this first order approximation ignores the convexity correction.

\subsubsection{Second order approximation}
A more accurate approximation can be derived by expanding up to order two. That is,
\begin{align}
V^2_{0,T} (t) T &\approx  (I^H \sqrt\tau)^2 + \frac{1}{2}  \frac{ \partial^2 (I^H_- \sqrt\tau)^2 }{\partial d^2_-} \int_{-\infty}^\infty  d^2_- N'(d_- ) \, d d_- \nonumber \\
&=  (I^H_- \sqrt\tau)^2 + \frac{1}{2}  \frac{ \partial^2 (I^H_- \sqrt\tau)^2 }{\partial d^2_-}  
\end{align}
This formula gives an intuitive interpretation of the convexity correction, namely as the convexity of the historical adjusted implied variance at the zero vanna strike:
\begin{equation}\label{convexity}
V^2_{0,T} (t) T - \mathcal{V}^2_{0,T} (t) T \approx V^2_{0,T} (t) T -  (I^H_- \sqrt\tau)^2 \approx \frac{1}{2}  \frac{ \partial^2 (I^H_- \sqrt\tau)^2 }{\partial d^2_-}
\end{equation}

\smallskip
To find the hedge ratio, note that
\begin{align}
dV^2_{0,T} (t) T -  D((I^H_- \sqrt\tau)^2) &= dV^2_{0,T} (t) T -  2 I^H_- \sqrt\tau \, D (I^H_- \sqrt\tau) - (D (I^H_- \sqrt\tau))^2 \nonumber \\
&\approx \frac{1}{2} D \left(  \frac{ \partial^2 (I^H_- \sqrt\tau)^2 }{\partial d^2_-} \right) \nonumber \\
&\approx  \frac{ \partial^2 (I^H_- \sqrt\tau)}{\partial d_-^2} \, D (I^H_- \sqrt\tau) + D (\cdots)
\end{align}
As $E (dV^2_{0,T} (t) T) = 0$ and $E (D (I^H_- \sqrt\tau)) \approx 0$ the terms not involving $dV^2_{0,T} (t) T $ and $D (I^H_- \sqrt\tau)$ will approximately cancel out. We are therefore left with
\begin{equation}
dV^2_{0,T} (t) T -  2 I^H_- \sqrt\tau \, D (I^H_- \sqrt\tau) \approx  \frac{ \partial^2 (I^H_- \sqrt\tau)}{\partial d_-^2} \, D (I^H_- \sqrt\tau) 
\end{equation}
Using \eqref{dvs=DI} we arrive at
\begin{equation}\label{secondorder}
d \mathcal{V} _{0,T} (t) \approx \frac{1} {2 I^H_- \sqrt\tau +  \frac{ \partial^2 (I^H_- \sqrt\tau) }{\partial d^2_-} } \, d V^2_{0,T} (t)
\end{equation}

\newpage
\bibliography{vshedging}
\bibliographystyle{abbrvnat}

\newpage

\appendix
\section{The Black-Merton-Scholes case}
\noindent
It is clear that equation \eqref{CBSH} is the key to pricing (seasoned) volatility swaps. In general it cannot be solved analytically and although approximations are possible, we would like to limit the number of approximations made in the pricing. Numerical integration is therefore preferable to analytical approximations when evaluating \eqref{CBSH}. In this section we will demonstrate  that when we limit ourselves to a Black-Scholes world with deterministic term structure an analytical solution can be obtained and we recover the exact price of a seasoned volatility swap.

\smallskip
In what follows we will need to evaluate a double integral involving the normal probability density function and its cumulative distribution function. Let
\begin{align}
\phi(x) &= \frac{1}{\sqrt{2\pi}} \exp\left\{-\frac{1}{2}x^2\right\} \\
\Phi(x) &= \int_{-\infty}^x \phi(y) dy
\end{align}
Then, for constants $a$ and $b$, it can be shown that
\begin{equation}
\label{gaussint}
\int_{-\infty}^\infty \Phi (a + bx) \phi(x) dx = \Phi \left( \frac{a}{\sqrt{1 + b^2}} \right)
\end{equation}
For the above and other Gaussian integrals we refer the reader to \cite{Owen}.

\smallskip
In a Black-Scholes world with deterministic term structure of volatility, the implied volatilities $I$ and those of the adjusted process $I^H$ will not depend on the strike of the option. Instead of trying to solve \eqref{CBSH} directly, we make the observation that the zero vanna implied volatility is equivalently characterized by the point where the price of a binary option is one-half (in the presence of a skew this is not the case anymore). Thus, for the option on the adjusted process we must have
\begin{equation}
- \left( \frac{ \partial C(S, K, I^H) }{\partial K} \right)_{I^H = I^H_- (t), K= K_-} = \frac{1}{2}
\end{equation}
Now we can differentiate the right-hand side of \eqref{CBSH} and find the strike $K$ and corresponding implied volatility $I$ such that the integral is equal to one-half:
\begin{equation}
- \frac{\partial}{\partial K} \int_0^\infty h C^{BS}(S,K/h,I) q(h) dh = \frac{1}{2}
\end{equation}
Introduce a new variable $y = \ln h$, then
\begin{equation}
- \frac{\partial}{\partial K} \int_0^\infty h C^{BS}(S,K/h,I) q(h) dh = \int_{ -\infty}^\infty \Phi (a + by) \frac{1}{\nu \sqrt{2\pi}} \exp \left\{  - \frac{1}{2} \left(\frac{ y - \mu}{\nu} \right)^2 \right\} dy
\end{equation}
where 
\begin{equation}
a = \frac{\ln S/K - \frac{1}{2} I^2\tau}{I \sqrt{\tau}}, \, b = \frac{1}{I \sqrt{\tau}}
\end{equation}
Define now the variable $y' = (y - \mu)/\nu$ and we obtain
\begin{align}
\int_{ -\infty}^\infty \Phi (a + by) \frac{1}{\nu \sqrt{2\pi}} \exp \left\{  - \frac{1}{2} \left(\frac{ y - \mu}{\nu} \right)^2 \right\} dy &= \int_{ -\infty}^\infty \Phi (a + \mu b + \nu b y') \phi(y') dy' \nonumber \\
&= \Phi \left( \frac{a + \mu b}{\sqrt{1 + \nu^2 b^2}} \right)
\end{align}
where we have used \eqref{gaussint}. This immediately gives the strike  where the vanna of the option on the adjusted process $S^H$ is zero:
\begin{equation}
\Phi \left( \frac{a + \mu b}{\sqrt{1 + \nu^2 b^2}} \right) = \frac{1}{2} \iff a = -\mu b
\end{equation}
Since $\mu = - \frac{1}{2} c^2 \tau$, the above condition gives
\begin{equation}
\ln S/K_- - \frac{1}{2} I^2\tau = \frac{1}{2} c^2 \tau
\end{equation}
which means that the zero vanna strike for the option on the adjusted process is
\begin{equation}
\ln S/K_- = \frac{1}{2} \left( c^2 + I^2 \right) \tau
\end{equation}
The zero vanna adjusted implied volatility $I^H_-$ satisfies
\begin{equation}
\ln S/K_- = \frac{1}{2}(I^H_-)^2 \tau
\end{equation}
and so,
\begin{equation}
I^H_-  = \sqrt{ c^2 + I^2}
\end{equation}
In the Black-Scholes model with deterministic volatility term structure $\sigma$,
\begin{equation}
c^2 = \frac{1}{\tau} \int_0^t \sigma^2 \, du, \, \, I^2 = \frac{1}{\tau} \int_t^T\sigma^2\, du
\end{equation}
and so \eqref{priceresult} gives us
\begin{equation}
\label{resultBS}
\mathcal{V}_{0,T}(t) \approx I^H_-  \sqrt{\frac{\tau}{T}} = \sqrt{ \frac{1}{T} \int_0^t \sigma^2 du + \frac{1}{T} \int_t^T \sigma^2du }
\end{equation}
but of course the approximate result is in fact exact in the Black-Scholes case.

\smallskip
We could have written down equation \eqref{resultBS} without carrying out the previous calculations since in a Black-Scholes world there is no uncertainty about realized volatility. Nevertheless, the fact that we arrive at the result using our method which is applicable to general stochastic volatility models, shows that equation \eqref{priceresult} is correct.

\end{document}